# Cr doped III-V nitrides: potential candidates for spintronics

B. Amin[1], S. Arif[1], Iftikhar Ahmad[1,*], M. Maqbool[2], R. Ahmad[3], S. Goumri- Said[4], K. Prisbrey[5]

1. Materials Modeling Lab., Department of Physics, Hazara University, Mansehra, Pakistan
2. Department of Physics and Astronomy, Ball State University, Indiana, USA
3. Department of Chemistry, Hazara University, Mansehra, Pakistan
4. Physical Sciences and Engineering Division, Ibn Sina Building, King Abdullah University of Science and Technology (KAUST), Box 4700, Thuwal 23955-6900, Saudi Arabia
5. Department of Metallurgy, University of Utah, Salt Lake City, USA

\* *Corresponding Author:*

Address:   Chair, Department of Physics and Astronomy, Hazara University, Mansehra, Pakistan

Email:     ahma5532@gmail.com

Phone:     (092)332-906-7866

**ABSTRACT**

Studies of Cr-doped III-V nitrides, dilute magnetic alloys, in the zinc-blende crystal structure are presented. The objective of the work is to investigate half-metallicity in $Al_{0.75}Cr_{0.25}N$, $Ga_{0.75}Cr_{0.25}N$ and $In_{0.75}Cr_{0.25}N$ for their possible application in the spin based electronic devices. The calculated spin polarized band structures, electronic properties and magnetic properties of these compounds reveal that $Al_{0.75}Cr_{0.25}N$ and $Ga_{0.75}Cr_{0.25}N$ are half-metallic dilute magnetic semiconductors while $In_{0.75}Cr_{0.25}N$ is metallic in nature. The present theoretical predictions provide evidence that some Cr doped III-V nitrides can be used in spintronics devices.

I. INTRODUCTION

Half-metallic diluted magnetic semiconductors (DMS) have attracted enormous attention due to their interesting physical properties and possible applications in the spin based devices[1-5]. These materials are alloys of semiconductors in which some cations are substituted by transition metal (TM) ions while the crystal field structure of the host material is maintained. They show band-gap at the Fermi level for one spin-state [6]. Half-metallic DMS are used for injecting spin polarized carrier into semiconductors and spin valves[7].

The III-nitride semiconductors are extensively studied for their optical and spin dependent electron transport properties [8-11]. Although, most of the III-nitride semiconductors are grown in the wurtzite structure but their zinc-blende structure (meta-stable) with large optical gain and low threshold current density has also been predicted[12-15]. The successful growth of the III-nitrides (AlN, GaN and InN) with the zinc-blende structure on GaAs substrate by molecular beam epitaxial technique has been reported[16, 17].

Half-metallicity is theoretically predicted in $Ga_{1-x}Mn_xN$ with 6.5% and 12.5 % of Mn concentrations [18-20] and also in $Al_{0.97}Mn_{0.03}N$[14]. It has been experimentally observed that the Curie temperature of AlCrN (2% Cr) is higher than AlMnN (2% Mn)[9]. Wu and Liu predicted 600 K Curie temperature for AlCrN with $0.05 \leq x \leq 0.15$ and experimentally found that Cr is magnetically active at 3% doping in GaN while 7% in AlN [21, 22]. The interesting physical properties of half-metallic dilute magnetic semiconductors and their possible potential applications in the miniaturization of the electronic devices[23] motivated us to investigate Cr doped III-V nitride semiconductors in the zinc-blende phase.

## II. Method of calculations

Kohn, Hohenberg and Sham proposed density functional theory (DFT), the most reliable theoretical tool, for the probing of the structural, electronic, optical and magnetic properties of metals, semi-metals, half-metals, semiconductors, insulators and superconductors. They used electrons density instead of electrons wave functions for the calculation of different physical properties of solids. The Kohn-Sham equations in atomic units are[24]:

$$[-\frac{1}{2}\nabla^2 + V_{ext}(\vec{r}) + V_C[\rho(\vec{r})] + V_{xc}[\rho(\vec{r})]]\phi_i(\vec{r}) = \varepsilon_i \phi_i(\vec{r}) \qquad (1)$$

the first term on the left represents the kinetic energy operator, the second the external potential from the nuclei, the third is the Coulomb potential and the fourth one is the exchange correlation potential. The Kohn-Sham equations are solved iteratively till self consistency is achieved. In the present work, the Kohn-Sham equations are solved using the full potential linearized augmented plane-wave (FP-LAPW) method[25] with the Generalized Gradient Approximation (GGA)[26]. In the generalized gradient scheme, the exchange-correlation energy ($E_{xc}$) is a functional of the local electron spin densities $\rho(r)$ and their gradients[24]:

$$E_{xc}^{GGA}(\rho_\uparrow, \rho_\downarrow) = \int \varepsilon_{xc}(\rho_\uparrow(\vec{r}), \rho_\downarrow(\vec{r}), \nabla\rho_\uparrow(r), \nabla\rho_\downarrow(r))\rho(\vec{r})d^3r \qquad (2)$$

where $\rho_\uparrow$ and $\rho_\downarrow$ are densities for spin up and spin down electrons and $\varepsilon_{xc}$ is exchange correlation energy per particle. Details of the spin-polarized FP-LAPW calculations, formulas and the wien2k software used in this work are reported by Schwarz and Blaha[27] and Blaha et al.[28].

In the full potential scheme; radial functions times wave function, potential and charge density are expanded into two different basis sets. Within each atomic sphere the wave function is expanded in spherical harmonics while in the interstitial region it is expanded in a plane wave basis. In the same manner the potential is expanded as[29]:

$$V(r) = \begin{cases} \sum_{lm} V_{lm}(r) Y_{lm}(\hat{r}) \dots\dots\dots\dots\dots\dots(a) \\ \sum_{K} V_{K} e^{iKr} \dots\dots\dots\dots\dots\dots\dots(b) \end{cases} \quad (3)$$

where, Eq. 3(a) is for the inside of the sphere and Eq. 3(b) is for the interstitial sites. Inside the sphere the wave function is expanded in terms of spherical harmonics up to $l = 9$. The potential is spherically symmetric within the muffin-tin sphere while outside the sphere it is constant. The core electrons are treated fully relativistically while the valence electrons are treated semi-relativistically. In order to ensure that no electron leakage is taking place semi-core states are included so that accurate results are achieved; 56-k points are used and $R_{MT} \times K_{max} = 8.00$ determines the plane wave basis functions.

## III. RESULTS AND DISCUSSION

### A. Structural Properties

To calculate the structural properties of $M_{0.75}Cr_{0.25}N$ (M=Al,Ga,In) compounds, the supercell volume of each material is optimized by using the Birch-Morgan's equation of state[30]. The relaxed structural parameters, lattice constants ($a$) and bulk moduli (B), are evaluated at the optimized volumes. The calculated results are presented in Table 1. The data presented in the table shows a small difference between our results and those reported by other calculations. A small difference in the results is due to the fact that, we used generalized gradient approximation (GGA) and in the other calculations they used local density approximation (LDA). But it is a well established fact that GGA calculations are superior to LDA and hence the GGA results are closer to the experimental results[31, 32].

The atomic volumes of Al, Ga, In, N and Cr are 10, 11.81, 15.71, 13.65 and 7.23cm$^3$/mol respectively. It is clear from the table that the partly substitution of M (M=Al, Ga, N) by Cr atoms affects the lattice constants of the zinc blende MN crystals. The decrease in the lattice constants, when one M atom is replaced by Cr atom in the supercell of eight atoms, is due to the considerable difference in the atomic volumes of M and Cr. The change in the lattice constant of MN by the substitutional atom of different size also produces a small distortion in the unit cell. The results presented in the table show that the calculated lattice constant of $Al_{0.75}Cr_{0.25}N$ is smaller than $Ga_{0.75}Cr_{0.25}N$, while the lattice constant of the later one is smaller than $In_{0.75}Cr_{0.25}N$. The difference in the lattice constants are due to the dissimilarity in the atomic volumes of In, Ga and Al ($V_{In}>V_{Ga}>V_{Al}$). It can further be noted from the table that the bulk modulus increases with the decrease in the lattice constant. This increase can be related to the increase in the hardness of a material and hence $Al_{0.75} Cr_{0.25}N$ is the hardest of all the three. The results are in agreement with Ref. 33 and Ref. 34.

The magnetic nature of $M_{0.75} Cr_{0.25}N$ (M=Al,Ga,In) materials is investigated by the difference between the antiferromagnetic and ferromagnetic energies ($\Delta E=E_{AFM} -E_{FM}$). The calculations show that ferromagnetic nature is favorable for all the three compounds. The ferromagnetic nature of Cr doped III-V dilute magnetic semiconductors at low concentrations is also reported by Kaminska et al.[35] and Frazier et al.[36] in their experimental studies. Our theoretical calculation confirms the ferromagnetic nature of these compounds at higher concentrations.

## B. Electronic Band Structure and Density of States

Along high symmetry directions in the first Brillouin zones, the calculated spin polarized band structures for $M_{0.75}Cr_{0.25}N$ (M=Al, Ga, In) are shown in Figs. 1, 2 and 3. These figures show that the majority spin (↑) states contain more number of electrons than the corresponding minority spins (↓) states. The calculated bandgaps are compared with the other calculated results in Table 1. The bandgaps of pure AlN (5.4 eV), GaN (3.2) and InN (1.0) crystal decrease to 3.762 eV, 3.10 eV and 0.9 eV respectively, by Cr (25%) doping. The decrease in the band-gaps is due to the presence of local strains and local electric fields generated by Cr atoms.

Spin polarized band structures for $Al_{0.75}Cr_{0.25}N$ and $Ga_{0.75}Cr_{0.25}N$ are presented in Figs.1 and 2. It is revealed by these figures that for the minority spin-states (↓) the top of the valence bands and the bottom of the conduction bands occur at the Γ symmetry points. Hence, these compounds are wide bandgap semiconductors like AlN and GaN for minority spin- states (↓). The band structures also show that these compounds are metallic for the majority spin- states (↑). The overall spin dependent half-metallicity, conductor in one spin- state (↑) and semiconductor in the other spin state (↓), is confirmed for $Al_{0.75}Cr_{0.25}N$ and $Ga_{0.75}Cr_{0.25}N$ compounds. The spin dependent half-metallicity is also reported for $Al_{0.97}Mn_{0.03}N$[14], $Zn_{0.75}(Fe/Co/Ni)_{0.25}Se$ [33], CaC, SrC, BaC and MgC[37, 7] compounds. The nature of $In_{0.75}Cr_{0.25}N$ compound is different than the rest of the two because valence and conduction bands of this compound crosses each other for majority spin- state (↑), and the bottom of the conduction band crosses the Fermi level for minority spin-state (↓). Hence $In_{0.75}Cr_{0.25}N$ is metallic for both spin-channels. It can also be noted from the majority spin- states (↑) (presented in Figs. 1, 2 and 3) that conduction bands and valence bands states overlap and there are no forbidden energy gap at the Fermi level. These completely filled bands, in which conduction and valance band states overlap around the Fermi level are mainly due to the Cr-3d and N-2p states. While cutoffs at the Fermi level can be seen for the minority spin-states (↓) of $Al_{0.75}Cr_{0.25}N$ and $Ga_{0.75}Cr_{0.25}N$, and hence for both compounds forbidden energy gaps exist at the Fermi level. Adamowicz and Wierzbicki[38] suggest that only the minority spin (↓) active state cause's global magnetic moment. Half metallic energy gaps of $Al_{0.75}Cr_{0.25}N$ and $Ga_{0.75}Cr_{0.25}N$ prevent these materials from the activation of the majority spin-state (↑). These bands can be observed clearly in Figs. 4 (a) for $Al_{0.75}Cr_{0.25}N$. The data presented in Table 1 also reveals that the; band gap decreases with the increase in the lattice constant. These results are in agreement with the other experimental [39] and theoretical results[8, 35].

The atomic and orbital origin of different bands can be explained by the spin polarized total and partial density of states of $M_{0.75}Cr_{0.25}N$ (M=Al, In), shown in Figs. 4 and 5 ($Ga_{0.75}Cr_{0.25}N$ is not shown). It is clear from Fig. 4(a) that for the majority spin- state (↑) of $Al_{0.75}Cr_{0.25}N$, the region between -8.0 and -3.5 eV is mainly dominated by the Cr-3d state with a small contribution from Al-s state. The states above these states are predominantly Cr-3d and N-2p, which cross the Fermi level and hence the material show metallic nature. It is also clear from the figure of minority spin- states (↓) that those states which were crossing the Fermi level are pulled by the conduction band and hence a large bandgap is produced at the Fermi level.

The differentiation between majority (↑) and minority spins (↓) is mainly due to the hybridization of the Cr-3d and N-2p states. It is clear from Fig. 5 that $In_{0.75}Cr_{0.25}N$ is metallic for both spin- states.

The nature of attraction in these dilute magnetic semiconductors can be described by the spd exchange splitting. The spd exchange splitting is calculated by the band edge spin splitting of the conduction band minima and the valence bands maxima at the $\Gamma$ symmetry point (for spin-up and spin-down states)[,45-47]:

$$\Delta E_C = E^{\downarrow}_{CBMin} - E^{\uparrow}_{CBMin} \qquad (4)$$

$$\Delta E_V = E^{\downarrow}_{VBMax} - E^{\uparrow}_{VBMax} \qquad (5)$$

where $E^{\downarrow}_{CBMin}$ and $E^{\uparrow}_{CBMin}$ are the conduction bands minima for the spin down and spin up states (at Γ point), while $E^{\downarrow}_{VBMax}$ and $E^{\uparrow}_{VBMax}$ are the valance bands maxima for the spin down and spin up states (at Γ point). The calculated values of $E^{\downarrow}_{CBMin}$, $E^{\uparrow}_{CBMin}$, $E^{\downarrow}_{VBMax}$ and $E^{\uparrow}_{VBMax}$ are presented in Table 2.

The band structures of $Al_{0.75}Cr_{0.25}N$, $Ga_{0.75}Cr_{0.25}N$ and $In_{0.75}Cr_{0.25}N$ presented in Figs. 1, 2 and 3 are used to calculate $\Delta E_c$ and $\Delta E_v$. It is clear from the band structures that the minima of the conduction band for the same compound are different for spin up and spin down states, while similar is the case for the valance bands. The difference of the conduction bands minima in the spin up and spin down states ($\Delta E_c$) as well as the difference in the valance band maxima in the spin up and spin down states ($\Delta E_v$) are used to calculate spd splitting. The cause of the difference between spin up (↑) and spin down (↓) states is the hybridization of Cr-3s, Cr-3d and N-2p states. The spin up states of $Al_{0.75}Cr_{0.25}N$ and $Ga_{0.75}Cr_{0.25}N$ are metallic while the spins down states are insulator. Unlike these compounds $In_{0.75}Cr_{0.25}N$ is metallic for both spin- states.

The calculated values of $\Delta E_c$ and $\Delta E_v$, presented in Table 2 indicate more effective potentials for minority spin- states (↓) than the majority spin (↑)-states. The spd exchange splitting is maximum for $Al_{0.75}Cr_{0.25}N$ and minimum for $In_{0.75}Cr_{0.25}N$. The variation in the spd exchange splitting can be related to the difference in magnetizations and bandgaps of the materials. An interesting feature of these compounds is crystal field splitting. The p orbitals of the N atoms generate an electric field, which splits the 5-fold degenerate Cr-3d state into non-degenerate Cr-dγ ($e_g$-doubly degenerate) and Cr-dε ($t_{2g}$-triply degenerate) states. For $Al_{0.75}Cr_{0.25}N$ ($Ga_{0.75}Cr_{0.25}N$); Cr-dγ/Cr-dε is centered at -1.08/-0.3(-1.35/-0.36) eV with a band width of 0.90/1.45(0.98/1.6) eV and located at 2.1/2.97(1.37/2.47) eV from the valence band. These parameters show that the wave function hybridizes very little for $e_g$ and strongly for $t_{2g}$ states with the p state of N, produce bonding and anti-bonding hybrids. The bonding hybrides lie within the valence band and the anti-bonding lie in the gap; which shifts the valence band to higher energies than $e_g$ state. Figs. 4 and 5 also provide information about the splitting energy of the two non-degenerate states ($\Delta E_{crystal} = E_{dε} - E_{dγ}$). The magnitude of splitting for $Al_{0.75}Cr_{0.25}N$, $Ga_{0.75}Cr_{0.25}N$ and $In_{0.75}Cr_{0.25}N$ are 1.68, 1.52 and 1.28 eV for majority spins (↑) only. The results also reveal that the repulsion between bonding and anti-bonding is largest for $Al_{0.75}Cr_{0.25}N$ and smallest for $In_{0.75}Cr_{0.25}N$. This is due to the fact that AlN has the smallest ionicity and high covalency among the three.

## C. Magnetic Properties

The calculated total and local magnetic moments for $M_{0.75}Cr_{0.25}N$ (M=Al, Ga, In) are presented in Table 2. The main source of magnetization in these materials is the unfilled Cr-3d states. It is clear from the table that the total magnetic moment for $Al_{0.75}Cr_{0.25}N$ is larger than $Ga_{0.75}Cr_{0.25}N$, while the later one is larger than $In_{0.75}Cr_{0.25}N$. On the basis of above discussion it is confirmed that the magnetic coupling strength is inversely related to the lattice constant [40].

GaMnAs is a zinc-blende crystal [41] but Yu et al. [42] and Masek and Maca [43] are of the opinion that, it can not be exactly described as a GaAs structure with all Ga substituted by Mn, but some of the Mn resides on the interstitial sites. Similarly in the zincblende $M_{0.75}Cr_{0.25}N$ (M=Al, Ga, In) crystals, some of the Cr atoms occupy interstitial sites. The Cr substituted M (M=Al, Ga, In) atoms are called Cr- substitutional sites and those (Cr) that occupy the interstitial sites are called Cr-interstitials. The net magnetic moment of the unit cell of $M_{0.75}Cr_{0.25}N$ (M=Al, Ga, In) is the result of the contributions of M (M=Al, Ga, In), Cr-substitutionals anti-site (N) and Cr-interstitials. The origin of the Cr magnetic moment can be related to the partially filled $t_{2g}$ level in the 3d state [21]. It can be noted from Table 2 that the local magnetic moment of Cr reduces from its free space charge value $3\mu_B$ to 2.26128 $\mu_B$ in $Al_{0.75}Cr_{0.25}N$, $3\mu_B$ to 2.47372 $\mu_B$ in $Ga_{0.75}Cr_{0.25}N$ and $3\mu_B$ to 2.79737 $\mu_B$ in $In_{0.75}Cr_{0.25}N$. This decrease in the magnetization of Cr is due to the spd hybridization, which induces small local magnetic moments on M, N anti-sites and interstitial sites. It is also evident from the table that the magnetic moment of the Cr-substitutional sites increases gradually, 2.26128 $\mu_B$ ($Al_{0.75}Cr_{0.25}N$) → 2.47372 $\mu_B$ ($Ga_{0.75}Cr_{0.25}N$) →2.79737 $\mu_B$ ($In_{0.75}Cr_{0.25}N$). The increase in the magnetic moment of Cr is due to the decrease in the magnetic moment at the Cr-interstitial site of the corresponding material. The negative signs of the local magnetic moments of the N anti-sites for $Ga_{0.75}Cr_{0.25}N$ and $In_{0.75}Cr_{0.25}N$ demonstrate that the induced magnetic polarization at N atoms is anti-parallel to Cr atom. These show the anti-ferromagnetic interaction between N-2p and Cr spins. It is also obvious from the table that the magnetic moment at the N anti-site defects increases from $Al_{0.75}Cr_{0.25}N$ to $Ga_{0.75}Cr_{0.25}N$ to $In_{0.75}Cr_{0.25}N$. Sadowsi et al.[44] reported in their experimental work that there should be a balance between TM interstitial and N anti-site defects, leading to the reduced density of one type upon the increase in the density of the other one.

The sum of the calculated magnetic moments of the individual sites for $Ga_{0.75}Cr_{0.25}N$ and $In_{0.75}Cr_{0.25}N$ is more than the net magnetic moments of the corresponding compounds (Table 2). The change in the net magnetic moment of a material is because of the negative p–d coupling between Cr-3d and N-2p states (in Cr-doped semiconductors). This negative coupling lowers the total energy which stabilizes the magnetic configuration of the compounds.

The net magnetic moments of $Al_{0.75}Cr_{0.25}N$ and $Ga_{0.75}Cr_{0.25}N$ are almost integer numbers, which is another evidence[7] of the half metallicity of these compounds.

The exchange constants $N_0\alpha$ and $N_0\beta$ show the contribution of the valence and conduction bands in the process of exchange and splitting, and can be related to the spin splitting at the gamma symmetry point of a band structure. The effect of the spd exchange on the band structure of a host semiconductor in the mean-field approximation is discussed in Ref. [46]. These parameters, $N_0\alpha$ and $N_0\beta$, are calculated directly from the conduction band-edge ($\Delta E_C = E^{\downarrow}_{CBMin} - E^{\uparrow}_{CBMin}$) spin-splittings and valence band-edge ($\Delta E_V = E^{\downarrow}_{VBMax} - E^{\uparrow}_{VBMax}$) spin-splittings of $M_{0.75}Cr_{0.25}N$ (M=Al, Ga, In), using the following relations[45-47]:

$$N_0\alpha = \frac{\Delta E_C}{x\langle M \rangle} \qquad (6)$$

$$N_0\beta = \frac{\Delta E_V}{x\langle M \rangle} \qquad (7)$$

where $\langle M \rangle$ is half of the spin magnetic moment of Cr ion [45-47] and $x$ is TM concentration. The calculated values of $\Delta E_c$, $\Delta E_v$, $N_0\alpha$ and $N_0\beta$ are presented in Table 2. The exchange constant varies from $Al_{0.75}Cr_{0.25}N$ to $Ga_{0.75}Cr_{0.25}N$ to $In_{0.75}Cr_{0.25}N$, which confirms the magnetic character of these materials. The data presented in the table also show large change in $N_0\beta$ and negligible change in $N_0\alpha$ for $Ga_{0.75}Cr_{0.25}N$ and $In_{0.75}Cr_{0.25}N$ compounds, which is in agreement with Ref. [47].

## IV. Conclusions

From the density functional calculations of the structural, electronic and magnetic properties of $M_{0.75}Cr_{0.25}N$ (M=Al, Ga, In), it is concluded that:

i. $Al_{0.75}Cr_{0.25}N$ and $Ga_{0.75}Cr_{0.25}N$ are metallic for the majority spin states while semiconducting for the minority spin state. Surprisingly, for both spins $In_{0.75}Cr_{0.25}N$ is metallic.

ii. $Al_{0.75}Cr_{0.25}N$ and $Ga_{0.75}Cr_{0.25}N$ show 100% spin polarization at the Fermi level, but no spin polarization at the Fermi level is observed for $In_{0.75}Cr_{0.25}N$.

iii. Cr doped AlN and Cr doped GaN are predicted to be potential materials for spintronic devices and need further theoretical and experimental investigations.

**Acknowledgments:**

Prof. Dr. Nazma Ikram, Ex. Director, National Center of Excellence in Solid State Physics is highly acknowledged for her valuable suggestions during this work.

**Figure Captions**

Fig. 1. Spin polarized band structures for $Al_{0.75}Cr_{0.25}N$ (a) Majority spin ($\uparrow$) (b) Minority spin ($\downarrow$)

Fig. 2. Spin polarized band structures for $Ga_{0.75}Cr_{0.25}N$ (a) Majority spin ($\uparrow$) (b) Minority spin ($\downarrow$)

Fig. 3. Spin polarized band structures for $In_{0.75}Cr_{0.25}N$ (a) Majority spin ($\uparrow$) (b) Minority spin ($\downarrow$)

Fig. 4(a). Total density of states (DOS) in B3 phase for $Al_{0.75}Cr_{0.25}N$

Fig. 4(b). Partial density of states (DOS) in B3 phase for $Al_{0.75}Cr_{0.25}N$

Fig. 5(a). Total density of states (DOS) in B3 phase for $In_{0.75}Cr_{0.25}N$

Fig. 5(b). Partial density of states (DOS) in B3 phase for $In_{0.75}Cr_{0.25}N$

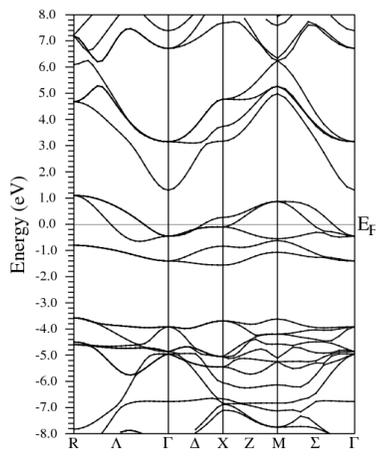 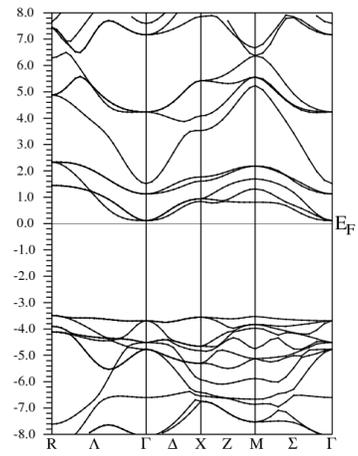

(a) (b)

Fig.1. Spin polarized band structures for $Al_{0.75}Cr_{0.25}N$ (a) Majority spin (↑) (b) Minority spin (↓)

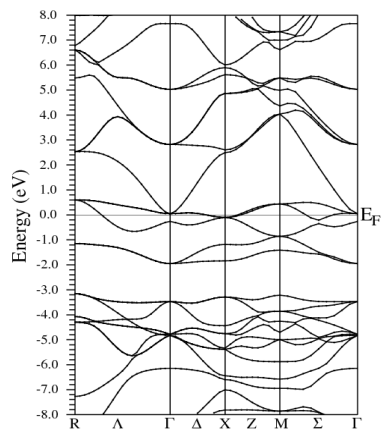 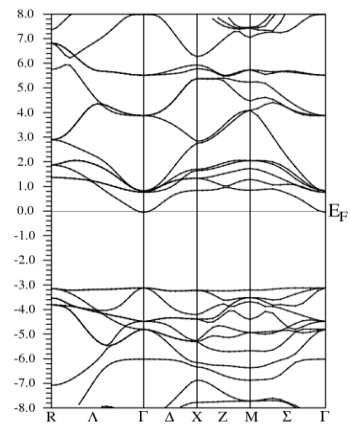

(a)                                          (b)

Fig. 2. Spin polarized band structures for $Ga_{0.75}Cr_{0.25}N$ (a) Majority spin (↑) (b) Minority spin (↓)

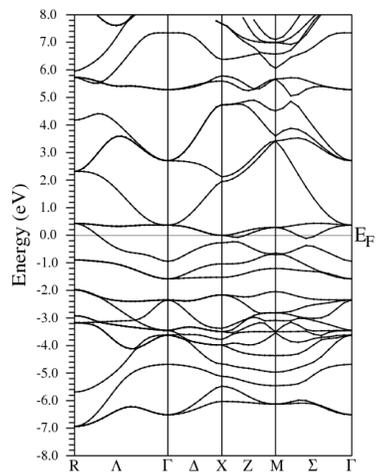 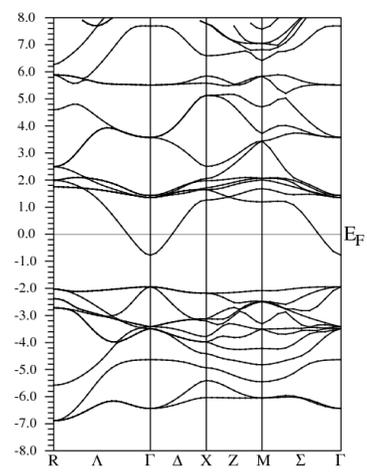

(a)                  (b)

Fig. 3. Spin polarized band structures for $In_{0.75}Cr_{0.25}N$ (a) Majority spin (↑) (b) Minority spin (↓)

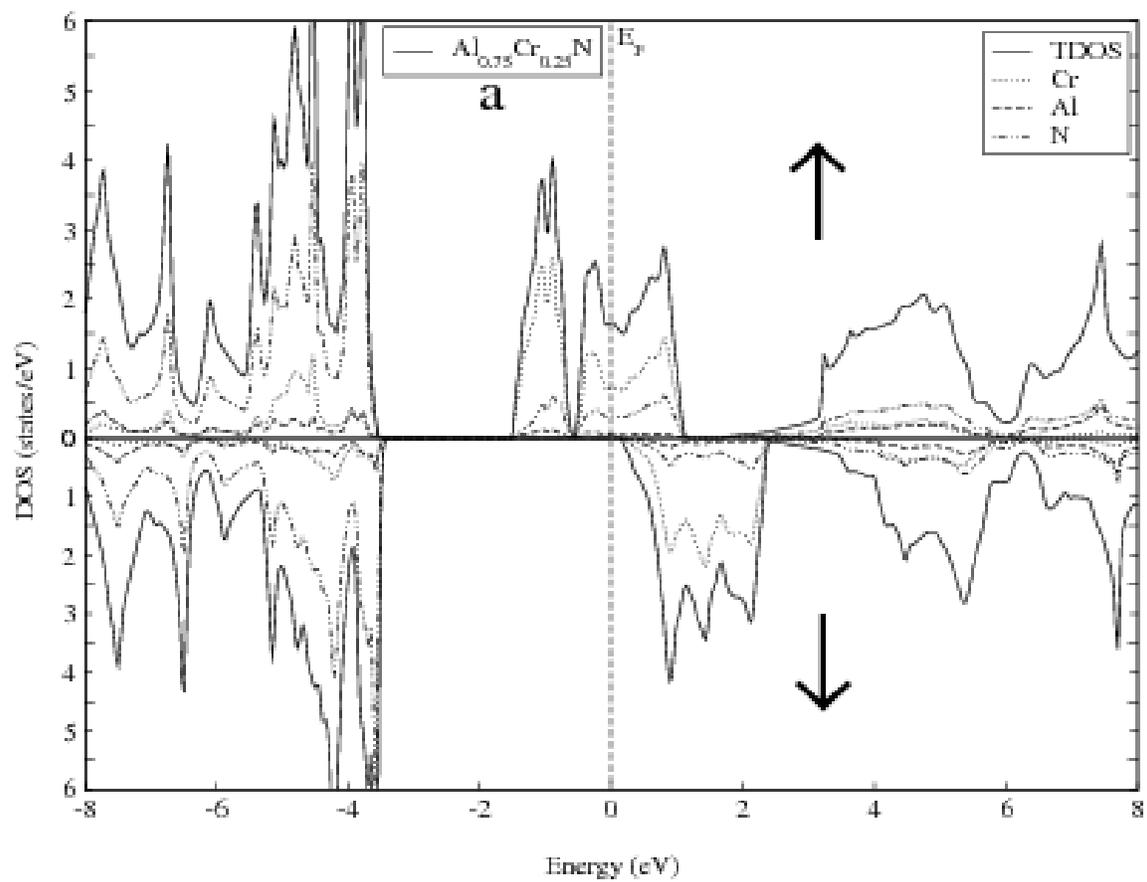

Fig 4(a). Total density of states (DOS) in B3 phase for $Al_{0.75}Cr_{0.25}N$.

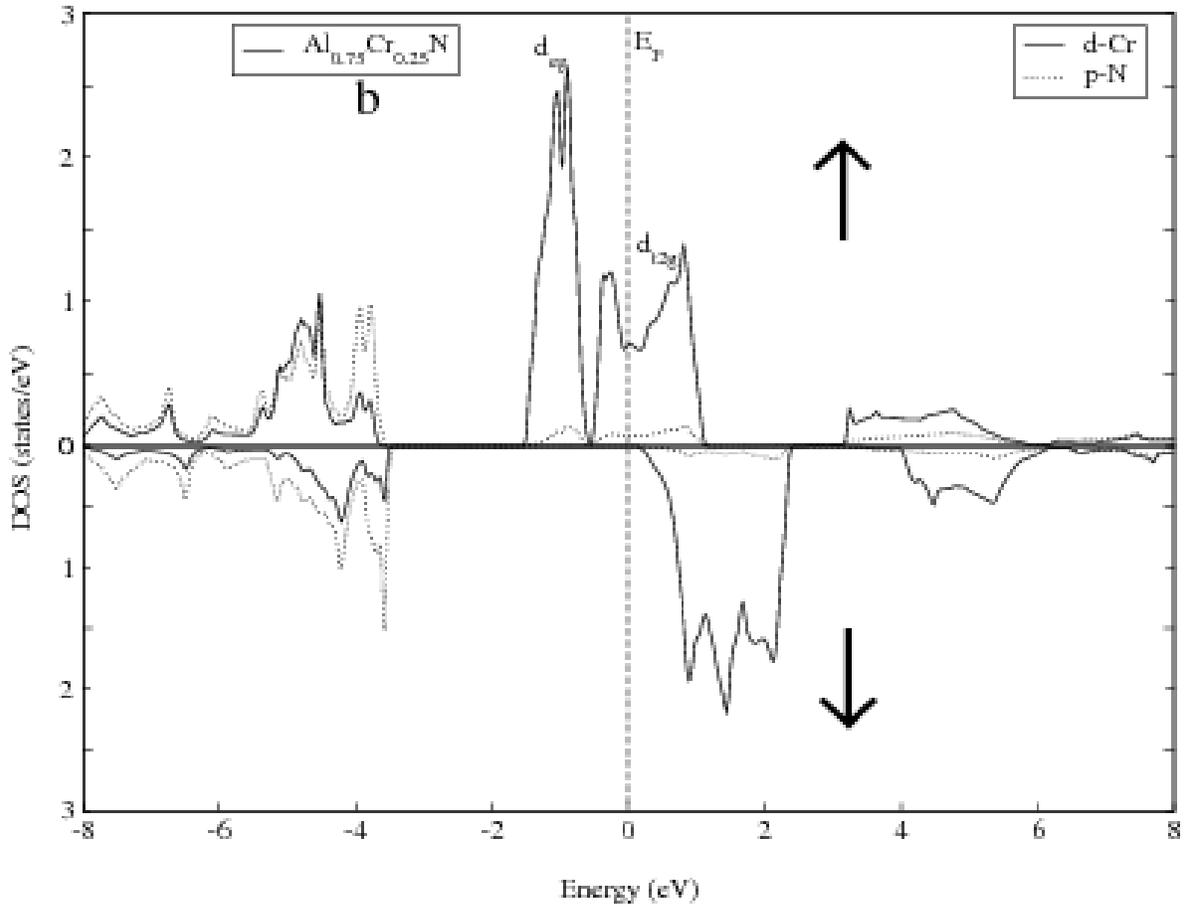

Fig. 4(b). Partial density of states (DOS) in B3 phase for $Al_{0.75}Cr_{0.25}N$

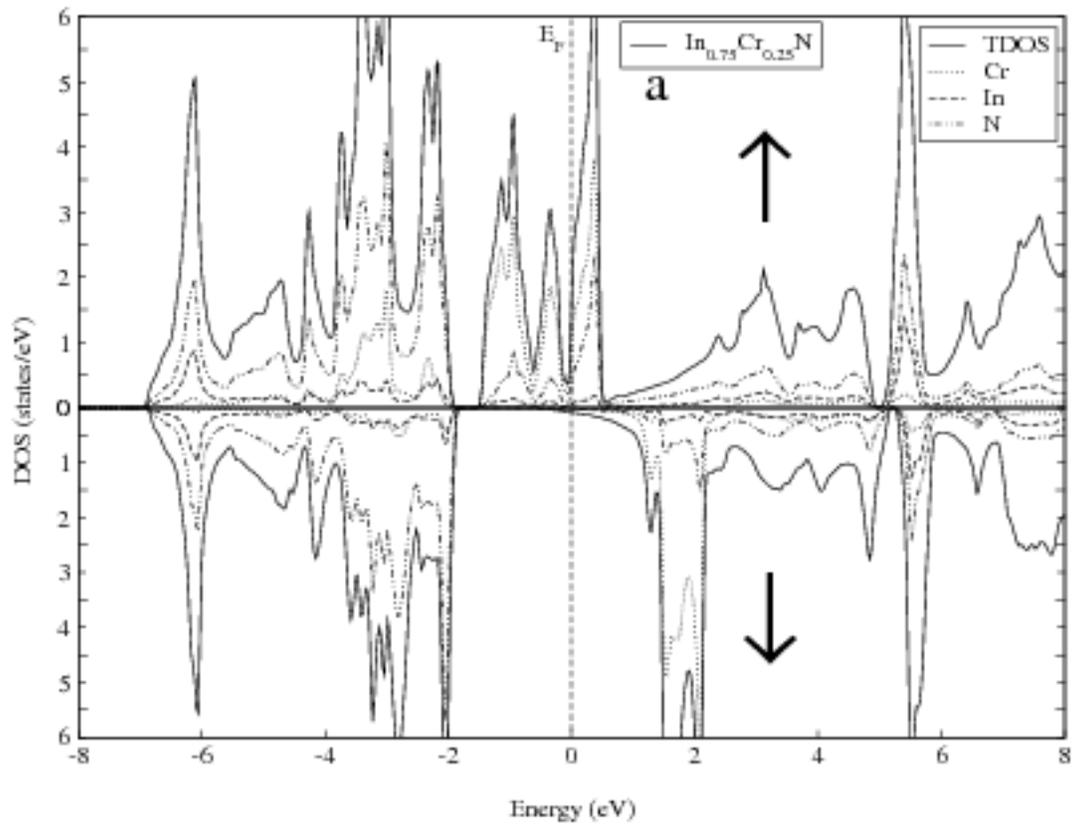

Fig. 5(a). Total density of states (DOS) in B3 phase for $In_{0.75}Cr_{0.25}N$

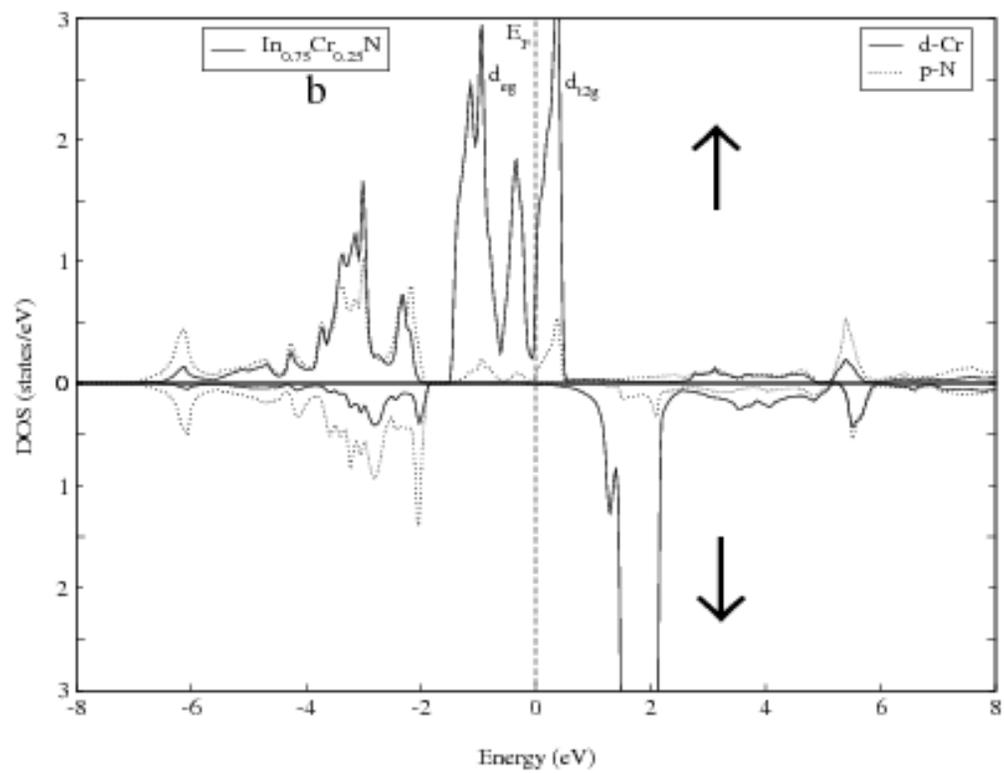

Fig. 5(b). Partial density of states (DOS) in B3 phase for $In_{0.75}Cr_{0.25}$ N

**Table.1.** Calculated lattice constant, bulk modulus and bandgap energy of (Al, Ga, In)$_{0.75}$Cr$_{0.25}$N

| Compounds | Calculations | Lattice constant (A$^0$) | Bulk modulus (Gpa) | Band-gap (eV) |
|---|---|---|---|---|
| AlN | This work | 4.40 | 192.6064 | 5.4 |
|  | Other Cal | 4.35[a] | 211.7[a] | 4.36[a] |
| Al$_{0.75}$Cr$_{0.25}$N | This work | 4.36 | 210.973 | 3.76 |
| GaN | This work | 4.58 | 165.753 | 3.2 |
|  | Other Cal | 4.50[b] | 202[a] | 3.07[b] |
| Ga$_{0.75}$Cr$_{0.25}$N | This work | 4.55 | 168.6807 | 3.10 |
| InN | This work | 5.01 | 119.8723 | 1.0 |
|  | Other Cal | 4.94[a] | 144.0[a] | 0.013[a] |
| In$_{0.75}$Cr$_{0.25}$N | This work | 4.95 | 129.7025 | 0.9 |

[a][47], [b][11]

**Table.2.** Total and local magnetic moments $\mu_B$ (in Bohr magneton) and exchange coupling constants for $(Al, Ga, In)_{0.75}Cr_{0.25}N$

| Sites ($\mu_\beta$) | $Al_{0.75}Cr_{0.25}N$ | $Ga_{0.75}Cr_{0.25}N$ | $In_{0.75}Cr_{0.25}N$ |
|---|---|---|---|
| $M^{Total}$ | 3.00084 | 3.00016 | 2.95674 |
| $M^{Cr}$ | 2.26128 | 2.47372 | 2.79737 |
| $M^{Al}$ | 0.02392 | --------- | ---------- |
| $M^{Ga}$ | --------- | 0.02211 | ---------- |
| $M^{In}$ | --------- | --------- | 0.01154 |
| $M^{N}$ | 0.01847 | -0.02728 | -0.07749 |
| $M^{Inter}$ | 0.67844 | 0.56606 | 0.43246 |
| $E_v^\uparrow$ | 0.000 | 0.00 | 0.00 |
| $E_v^\downarrow$ | -3.60 | -3.10 | -2.00 |
| $E_c^\uparrow$ | 0.00 | 0.00 | 0.00 |
| $E_c^\downarrow$ | 0.162 | 0.00 | 0.00 |
| $\Delta E_c$ | 0.162 | 0.00 | 0.00 |
| $\Delta E_v$ | -3.60 | -3.10 | -2.00 |
| $N_0\alpha$ | 0.216 | 0.00 | 0.00 |
| $N_0\beta$ | -9.60 | -8.20 | -5.33 |